 \documentclass[prb,aps,showpacs,twocolumn,a4paper,floatfix]{revtex4}
\usepackage{epsfig}
\usepackage{dcolumn}
\begin{document}
\title{Lattice dynamics of mixed semiconductors (Be,Zn)Se
from first-principles calculations}
\author{A.~V.~Postnikov}
\email{apostnik@uos.de}
\affiliation{Institute of Metal Physics, S. Kowalewskoj 18,
Yekaterinburg 620219, Russia, and
Universit\"at Osnabr\"uck --
Fachbereich Physik, D-49069 Osnabr\"uck, Germany}
\author{Olivier Pag\`es, Joseph Hugel}
\affiliation{Universit\'e de Metz - Institut de Physique,
1 Bd Arago, F-57078 Metz cedex 3, France}
\date{\today}
\begin{abstract}
Vibration properties of Zn$_{1-x}$Be$_x$Se, a mixed II-VI semiconductor 
characterized by a high contrast in elastic properties of its pure 
constituents, ZnSe and BeSe, are simulated by first-principles calculations 
of electronic structure, lattice relaxation and frozen phonons. 
The calculations within the local density approximation has been done 
with the {\sc Siesta} method, using norm-conserving pseudopotentials and 
localized basis functions; the benchmark calculations for pure endsystems 
were moreover done also by all-electron WIEN2k code.
An immediate motivation for the study was to analyze, at the microscopic level,
the appearance of anomalous phonon modes early detected in Raman spectra 
in the intermediate region (20 to 80\%) of ZnBe concentration. 
This was early discussed on the basis of a percolation phenomenon,
i.e., the result of the formation of wall-to-wall --Be--Se-- chains 
throughout the crystal. The presence of such chains was explicitly allowed 
in our simulation and indeed brought about a softening and splitting off 
of particular modes, in accordance with experimental observation, due to 
a relative elongation of Be--Se bonds along the chain as compared to those 
involving isolated Be atoms. The variation of force constants with interatomic 
distances shows common trends in relative independence on the short-range 
order.
\end{abstract}
\maketitle
\section[#1]{Introduction}

(Be,Zn)Se is an example of mixed II-VI semiconductor system
whose electronic properties and, consequently, elastic characteristics
vary with concentration in a non-trivial way. A considerable effort 
has been spent on characterizing the optical band gap, which drops
slightly from intermediate concentrations 
and undergoes a change from direct to indirect character underway
from ZnSe to BeSe \cite{PRB61-5332}.
Not less interesting possibilities open in the tuning of elastic properties,
because BeSe and ZnSe, similarly to some other easily miscible
II-VI constituents, possess quite different elastic constants, which have
to be accommodated in a solid solution. (Be,Zn)Se makes, actually, 
quite an extreme case of elastic contrast between constituents in a mixed 
system. This was argued to favor the appearance of anomalous lines in the
Raman spectra, which split off on the soft side of the BeSe-related TO peak 
in the $x\sim$0.2--0.7 concentration range of Be$_x$Zn$_{1-x}$Se crystals, 
as was detected by Pag\`es \emph{at al.} \cite{PRB65-035213}.
In order to explain this anomaly, Pag\`es \emph{et al.} \cite{PRB70-155319}
brought into discussion an idea of quasi-infinite chains, which are developing
in the softer matrix (ZnSe) as the BeSe concentration reaches 
the percolation threshold of $\sim$20\%.
However, the microscopic mechanism relating percolation with
vibration properties remained speculative. The present work's aim 
is to provide a possible complete and reliable first-principles description
of internal tensions, structural relaxation, and the impact 
of the latter on lattice-dynamical properties of a mixed Be$_x$Zn$_{1-x}$Se
alloy near the percolation threshold. A concise result of our simulation
has been reported earlier as a conference proceeding \cite{Dui03-ZBS}. 
Here we offer a detailed outline of results obtained for different 
supercells, with a more lengthy discussion.
In particular, we provide an analyzis of bond lengths distribution
for a number of compositions, discuss the change of force constants
related with that, and finally calulate phonon density of states and
provide its the wavevector-resolved decomposition.

The electronic properties of pure constituent compounds are well known;
they also were probed by \emph{ab initio} methods of respectful accuracy
in a number of earlier publications. 
Simulations of comparable accuracy on mixed alloys are seriously complicated
by the necessity to treat large supercell, with a practical loss of
any useful symmetry. A certain success has been achieved in molecular
dynamics (MD) simulations of an isostructural III-V mixed semiconductor
alloy, (Ga,In)As: Branicio \emph{et al.}\cite{APL82-1057} performed
parametrized MD calculations for supercells with up to 8000 atoms, and
\emph{ab initio} MD calculations on 216-atom supercells.
We are not aware of any elastic simulations
on mixed II-VI semiconductors but that by Tsai \emph{et al.} 
\cite{PRB65-235202}, who used multicenter MD method,
a formally \emph{ab initio} one but of inferior accuracy to the method
applied here. Comparing our results with those of Tsai \emph{et al.},
we find an overall good agreement, albeit with quantitative discrepancies
on some sensitive issues. A more important difference is, however, that
we study more attentively the limit of low Be concentration, particularly
the onset of percolation on the Be sublattice. Moreover, we
calculate phonon densities of states and discuss them in immediate
reference to experimentally obtained Raman spectra. 

The paper is organized as follows. 
In Sec.~\ref{sec:method} we outline our \emph{ab initio} calculation approach
to electronic structure and vibration properties.
Sec.~\ref{sec:pure}, containing the results for pure endsystems ZnSe and BeSe, 
is concise in view of numerous previous studies, but it is necessary here 
in order to assess the accuracy of the calculation method.
In Sec.~\ref{sec:bonds} we outline our results for lattice parameters
and equilibrium bond lengths in relaxed supercells, simulating a broad
concentration range.
The effect of bond lengths on the force constants between nearest
and next-nearest neighbors is discussed in Sec.~\ref{sec:forces}.
Finally, the manifestation of the force constants in phonon frequencies
is explained in Sec.~\ref{sec:phonons}, along with the discussion
on wavevector dependency and vibration patterns in different parts
of the phonon spectrum.

\section[#2]{Calculation scheme and setup}
\label{sec:method}
Dynamical properties are derived from the total energy or forces,
which are evaluated \emph{ab initio} in a sequence of 
density functional calculations. We applied the calculation method,
and computer code, {\sc Siesta}\cite{JPCM14-2745},
which incorporates norm-conserving pseudopotentials in combination
with atom-centered strictly confined numerical basis 
functions\cite{JPCM8-3859,PRB64-235111}.
The pseudopotentials were constructed along the Troullier--Martins
scheme\cite{PRB43-1993} for the following valence-states configurations:
Be$\,2s^2(2.0)$, Se$\,3d^{10}(1.2)\,4s^2(1.9)\,4p^4(2.0)$,
Zn$\,3d^{10}(1.09)\,4s^2(2.28)$. The numbers in brackets stand for
pseudoization radii of corresponding states in Bohr.
The basis set consisted of double-$\zeta$ functions with polarization
orbitals in all $l$ channels on all centers (Be$\,2s2p$, Zn$\,3d4s4p$,
Se$\,3d4s4p$), hence also for explicitly included semicore Se$3d$ states.
The details on the basis classification and testing in {\sc Siesta}
can be found in Refs.~\onlinecite{JPCM8-3859,PRB64-235111}.
Our basis generation followed the standard split schema
of the {\sc Siesta} code, with the energy shift parameter, responsible
for the localization of basis functions, equal to 20 mRy. This resulted
in basis functions with maximal extension of 3.17 {\AA} (Be),
3.12 {\AA} (Zn) and 2.82 {\AA} (Se), i.e. well beyond the nearest-neighbor
distances but short of direct overlap to next-nearest neighbors.
The full structure optimization 
(of cell parameters and internal coordinates) was performed in all
supercells prior to lattice dynamics calculations. In order to get reliable
forces on atoms it is essential to accurately perform spatial integration
of the (general-form) residual charge density over the unit cell.
To this end, we used the mesh cutoff parameter of 250 Ry, that generated
a real-space mesh with the step of about 0.1 {\AA} along each Cartesian
direction throughout the supercell. Within each mini-cell of this real-space
mesh, an averaging of integration results was done over four fcc-type sampling
points for better numerical stability. This resulted in the forces
summing up, over all atoms in the supercell, to zero within the accuracy of
$\pm 0.1$ eV/{\AA}, separately along each Cartesian direction.
The thus validated forces were used to calculate the dynamical matrix
elements, by introducing small (0.016 {\AA}) deviations of atoms
one by one from their equilibrium positions. The details of phonon calculations
are discussed in Sec.~\ref{sec:phonons}.

All our {\sc Siesta} calculations have been done in the
local density approximation (LDA). 
For the pure ZnSe and BeSe systems, moreover, we used the full-potential
augmented plane wave method (see, e.g., Ref.~\onlinecite{Singh_bookPWPP}),
as implemented in the WIEN2k code\cite{wien2k}. We apply here this all-electron
method of recognized accuracy for benchmark calculations of elastic properties.
We provide the WIEN2k results obtained both within the LDA and 
in the generalized gradient approximation (GGA, Ref.~\onlinecite{PRL77-3865}).

\begin {table*}
\caption{\label{tab:elastic}
Elastic properties of ZnSe and BeSe from experiments and
first-principles calculations.
}
\begin{ruledtabular}
\begin{tabular}{lccccccc}
 \rule[-2mm]{0mm}{7mm}
 & \multicolumn{3}{c}{ZnSe} && \multicolumn{3}{c}{BeSe} \\
 \cline{2-4} \cline{6-8}
 Method & $a$ ({\AA}) & $B$ (Kbar) & $\omega_{\mbox{\tiny TO}}$ (cm$^{-1}$) &&
          $a$ ({\AA}) & $B$ (Kbar) & $\omega_{\mbox{\tiny TO}}$ (cm$^{-1}$) \\
\hline 
WIEN2k (LDA)       & 5.568 &   714    & 198     && 5.084     & 920 & 498 \\
WIEN2k (GGA)       & 5.571 &   727    & 206     && 5.182     & 816 & 523 \\
{\sc Siesta} (LDA) & 5.587 &   758    & 203     && 5.078     & 912 & 500 \\
exp.
                   & 5.668\footnotemark[01] 
                           & 624\footnotemark[02]
			              & 207,\footnotemark[03]
			                205\footnotemark[04]
	          && 5.137\footnotemark[05]
		           & 920\footnotemark[05]
			              & 501\footnotemark[04] \\
other calc.
                   & 5.677,\footnotemark[06]  
                     5.638,\footnotemark[07]  
		           & 689,\footnotemark[06]
		             652,\footnotemark[07]
			              & 224\footnotemark[08] 
	          && 5.037\footnotemark[09] 
		           & 988\footnotemark[09]
			              & 547\footnotemark[10] \\
                   & 5.633,\footnotemark[08] 
                     5.636\footnotemark[11]  
		           & 811,\footnotemark[08]
		             649\footnotemark[11] \\
\end{tabular}
\end{ruledtabular}
\footnotetext[01]{Ref.~\onlinecite{LB17b}}
\footnotetext[02]{Ref.~\onlinecite{JAP41-2988}}
\footnotetext[03]{Ref.~\onlinecite{APL77-519}}
\footnotetext[04]{Ref.~\onlinecite{CrRT38-359}}
\footnotetext[05]{Ref.~\onlinecite{PRB52-7058}}
\footnotetext[06]{LDA, norm-conserving pseudopotential 
                  with Zn$3d$ as valence states, Ref.~\onlinecite{PRB51-10610}}
\footnotetext[07]{LDA, norm-conserving pseudopotential with Zn$3d$ and Se $3d$
                  as valence states, Ref.~\onlinecite{PRB52-1459}}
\footnotetext[08]{LDA, full-potential LMTO, Ref.~\onlinecite{PRB50-14881}}
\footnotetext[09]{LDA, norm-conserving pseudopotential, 
                  Ref.~\onlinecite{PRB54-11861}}
\footnotetext[10]{LDA, norm-conserving pseudopotential, 
                  Ref.~\onlinecite{PRB55-14043}}
\footnotetext[11]{full-potential LMTO, Ref.~\onlinecite{PRB64-035206}}
\end{table*}

\section[#3]{Endsystems, Z\lowercase{n}S\lowercase{e} 
and B\lowercase{e}S\lowercase{e} }
\label{sec:pure}

\begin{figure}[b]
\centerline{\epsfig{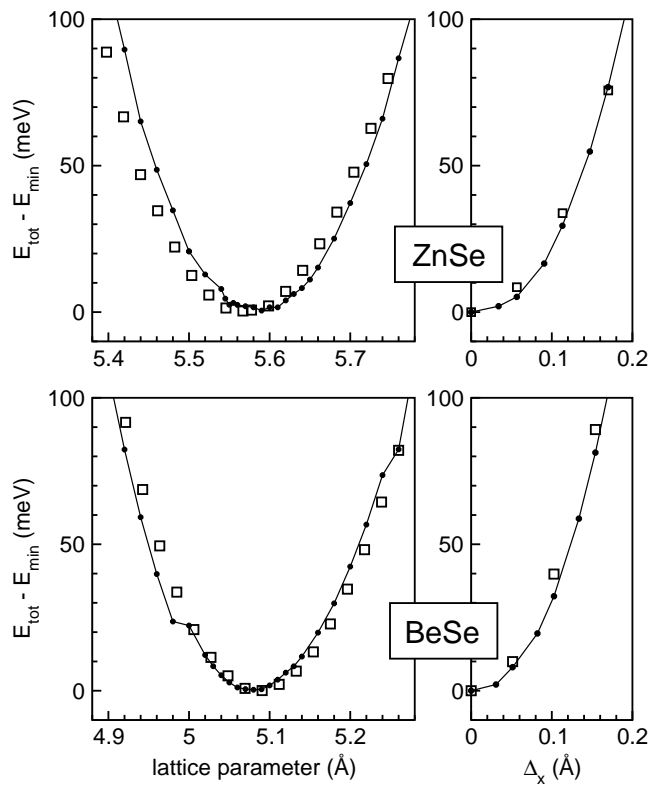}}
\caption{\label{fig:elastic}
The total energy vs. lattice constant curves (left) 
and the energy profiles corresponding to the $\Gamma$-TO phonon (right) 
as calculated for ZnSe and BeSe.
$\Delta_x$ indicates the magnitude of the cation (or anion)
displacement along [100] from the equilibrium.
Open squares: WIEN2k (LDA) results; connected black dots: {\sc Siesta}
(LDA) results.
}
\end{figure}
The electronic structure of zincblende-structure semiconductors
ZnSe and BeSe was extensively studied by both experimental means
and first-principle calculations. 
Band dispersions and partial densities of states have been calculated
in the LDA by Lee \emph{et al.}\cite{PRB52-1459} 
and Agrawal \emph{et al.}\cite{PRB50-14881} (for ZnSe, among other
Zn chalcogenides) and by Fleszar and Hanke\cite{PRB62-2466} (for BeSe,
among other Be chalcogenides). In the latter publication, also the
quasiparticle band structure has been calculated in the $GW$
approximation, and subsequently discussed in the context of LDA vs.
$GW$ and exact exchange for a number of $sp$ zincblende 
semiconductors\cite{PRB64-245204}. For ZnSe, the quasiparticle 
band structure has been reported by Luo \emph{et al.}\cite{PRB66-195215}.
Our calculations provide the band structures in agreement with 
good earlier LDA results, so we skip the discussion on this point.
Instead we turn to elastic properties and note that the optimized 
ground-state volume along with the bulk modulus have been reported 
in many calculations for ZnSe (see Table 1), and also
by Mu{\~n}oz \emph{et al.}\cite{PRB54-11861} and Gonz\'ales-D{\'{\i}}az
\emph{et al.}\cite{PRB55-14043} for BeSe. The latter calculations 
largely overestimate the stiffness of the BeSe crystal, probably due to 
the attribution of the Se$3d$ states to the core. As the treatment of Zn$3d$ 
(and also preferably Se$3d$) as valence states is now recognized as essential
in pseudopotential calculations for achieving good description of elastic 
properties of ZnSe, an overall agreement is established between 
state-of-art LDA calculations. 
Fig.~\ref{fig:elastic} shows the total energy profiles, as function
of uniform lattice scaling (left panels) and in its dependency 
on the off-center displacement (along any Cartesian direction) of one 
sublattice (anions or cations) relative to the other, i.e., the
zone-center TO mode (right panels).
The energy vs. lattice constant curves from all-electron WIEN2k calculations
and from {\sc Siesta} are almost identical. Small kinks in the {\sc Siesta}
total energy plot at some values of lattice constant are due to changes
of the real-space mesh for spatial integration (as the cell volume varies
but the mesh density is maintained about the same).
The energy profiles of the anion--cation displacement (TO phonon,
see right panels of Fig.~\ref{fig:elastic}) shows no substantial anharmonicity 
(deviations from parabolic behavior)
for both ZnSe and BeSe. The phonon frequencies in Table~\ref{tab:elastic}
were calculated from the second-order fit to these data.
Phonon frequencies in the following sections result from the diagonalization
of dynamical matrices, constructed from the forces induced on atoms 
by small Cartesian displacements. For pure constituents they agree very well
with the results of the second-order total energy fit.
It is noteworthy that a big difference in $\Gamma$-TO frequencies 
of ZnSe and BeSe is only in part due to a large differences
in the cation masses. The force constants are larger in BeSe,
as is well seen from Fig.~\ref{fig:elastic}, indicating higher stiffness 
of this material, also manifested in its larger bulk modulus.

Some our results listed in Table 1 are slightly refined as compared to
those reported earlier in Ref.~\onlinecite{Dui03-ZBS}, 
due to an use of a more recent realization of the all-electron method
(WIEN2k instead of WIEN97), and a more extended basis set in {\sc Siesta}.

In the discussion of phonon properties below we refer to dispersion curves
of ZnSe (obtained from neutron scattering and calculated in the rigid
ion model, Ref.~\onlinecite{PhL36A-376}) and BeSe
(calculated using self-consistent pseudopotential method and
linear response approach, Ref.~\onlinecite{BeSe-Tutuncu}).
A recent calculation\cite{PSSB229-563} of phonon dispersion for both systems 
based on a simple model with central and angular forces 
yields acceptable results for ZnSe but large difference 
from \emph{ab initio} phonon dispersion curves for BeSe. This could be 
a manifestation of more covalent character of bonding in BeSe, 
as discussed below.

\section[#4]{Bond lengths in mixed crystals}
\label{sec:bonds}
\begin{figure}
\centerline{\epsfig{figure=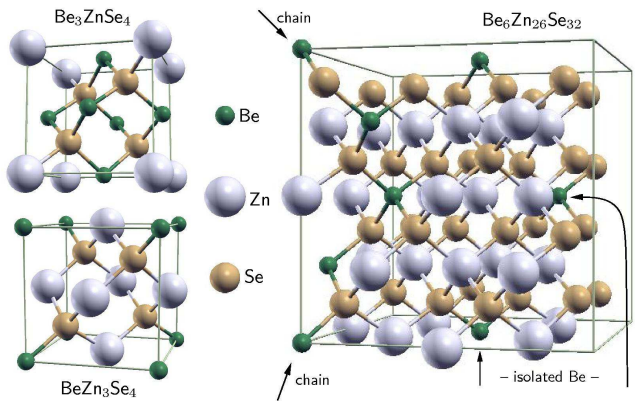,width=8.5cm}}
\caption{\label{fig:all_scells}
Cubic supercells used in the calculation. The continuous Be--Se chain
and isolated Be atoms are marked in the large supercell.
The figure was created with the XCrySDen software \cite{xcrysden}.
}
\end{figure}

In a sequence of supercell calculations which represent mixed Be--Zn
compositions, we begin by unconstrained relaxation of lattice vectors
and internal coordinates. 
Probably the simplest model of a mixed system with predominantly
either Be or Zn at cation sites is an ordered 8-atom superstructure with
the cubic primitive cell. Such supercells BeZn$_3$Se$_4$ and Be$_3$ZnSe$_4$ 
are shown in the left-hand side of Fig.~\ref{fig:all_scells}.
We emphasize that the BeZn$_3$Se$_4$ superstructure contains Be atoms
only in configuration which will be further referred to as ``isolated'' one;
i.e., each Se neighbor to a Be atom has only Zn ions to saturate
its other bonds. In the Be$_3$ZnSe$_4$ superstructure, on the contrary,
each Se ion has but a single Zn neighbor, whereas Be--Se bonds
form a continuous framework throughout the crystal.

In order to better understand the changes of the phonon spectra
in presumed dependency on the percolation setup, we performed 
the lattice dynamics simulation in a larger 2$\times$2$\times$2
(64 atoms) supercell. It contained 4 beryllium atoms in a continuous
--Be--Se-- chain transversing the crystal; moreover,
two other Be atoms were situated in ``isolated'' positions, with
only their fifth-nearest neighbors being of the Be type. This supercell 
is also shown in Fig.~\ref{fig:all_scells}.
When fully relaxed, it maintains a nearly cubic shape, 
with a tiny $z$-compression ($c/a$=0.998) due to the anisotropy of chains.
The nominal composition of Be is 18.75 at.\%, i.e. just below the theoretical
percolation threshold (0.198) on the fcc lattice\cite{Stauffer-book}; 
however, our structure includes percolation on the Be sublattice 
by construction. The advantage of our choice
of supercell is that, despite its relatively moderate size, it allows
to analyze local properties of areas with percolation (chained Be--Se bonds)
or without percolation (isolated Be--Se bonds only) on equal footing.
The results for this supercell will constitute the bulk of our discussion
on the lattice dynamics.

We note in passing that possible manifestation of percolation effects in mixed
semiconductor systems, notably at the above critical concentration and
especially in systems with large contrast in stiffness, has been pointed at
by Bellaiche \emph{et al.} already some time ago\cite{PRB54-17568}.

Finally, four supercells Be$_n$Zn$_{32-n}$Se$_{32}$ ($n$=1,\dots,4)
of the same size and cubic shape as the latest one simulate a gradual
aggregation of Be ions neighboring \emph{the same} anion site.

\begin{figure}[b]
\centerline{\epsfig{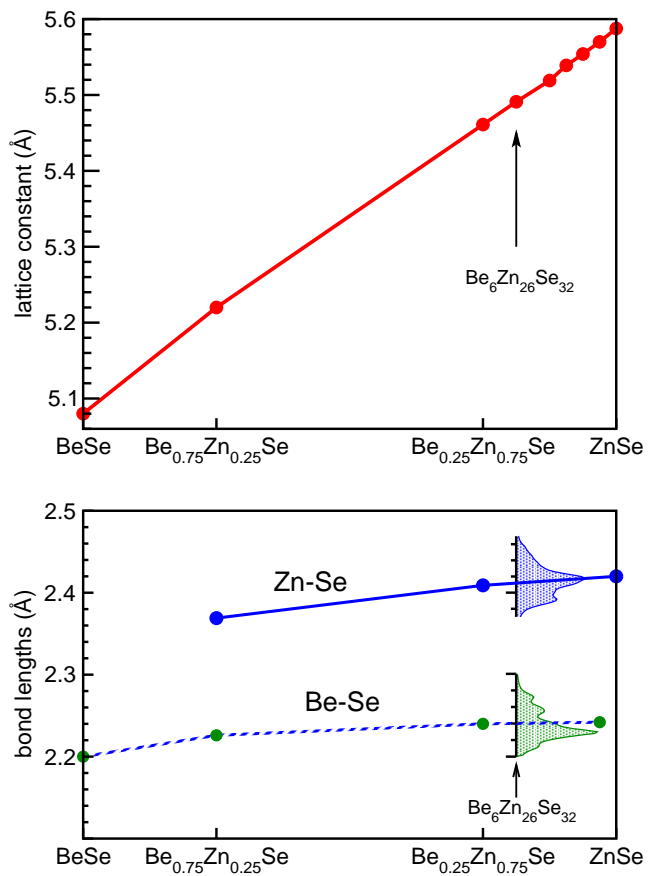}}
\caption{\label{fig:lattice}
Top panel: calculated equilibrium lattice constant in Be$_x$Zn$_{1-x}$Se
crystals of different concentration, represented by supercells
(see text for details).
Bottom panel: equilibrium Be--Se and Zn--Se bond lengths.
For $x$=0.19, the bond length values show a scattering due to the presence
of different local environments -- see details in the text
and in Fig.~\ref{fig:bonds}.
}
\end{figure}

Fig.~\ref{fig:lattice} shows the equilibrium lattice constant
(derived from the relaxed supercell volume)
and the mean cation--anion bond lengths. 
The mean lattice parameter exhibits a markedly linear dependence
on the concentration. At the same time, the Zn--Se and Be--Se bond lengths
tend to remain nearly constant throughout the whole concentration range.
A similar conclusion follows from the MD simulations for the (Ga,In)As system
by Branicio \emph{et al.}\cite{APL82-1057}, who also cite EXAFS data,
agreeing well with their results, and for the (Be,Zn)Se alloy
-- by Tsai \emph{et al.}\cite{PRB65-235202}, also based on MD simulations. 
However, the later publication reports a seemingly too large drop 
in the Be--Se distance when going from Zn$_{1/4}$Be$_{3/4}$Se to pure BeSe. 
It should be noted that Tsai \emph{et al.} found much smaller difference 
between the Zn--Se and Be--Se bond lengths than in our case, moreover
their lattice parameters for pure constituents considerably deviate 
from experiment, probably, due to insufficiency of the basis set
(and/or pseudopotential) they used.

\begin{figure}
\centerline{\epsfig{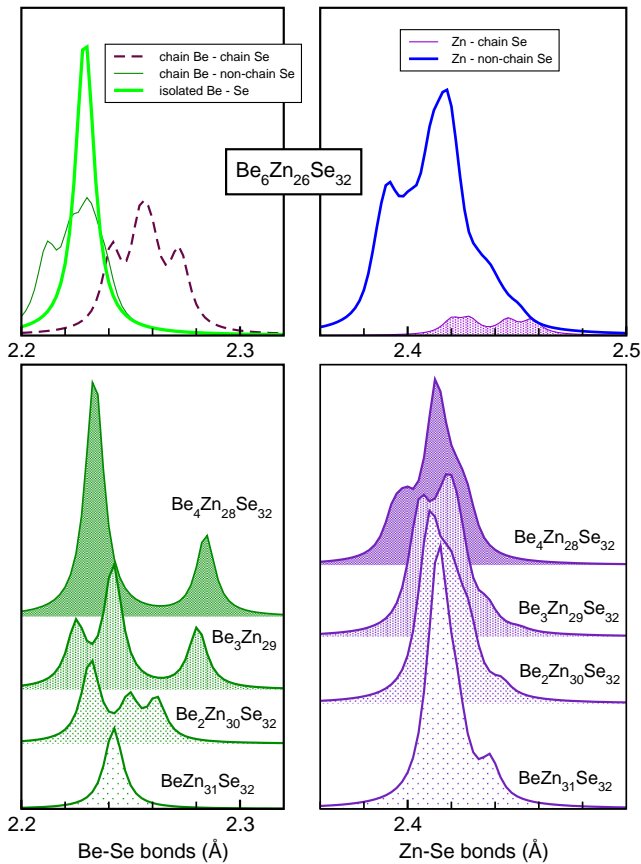}}
\caption{\label{fig:bonds}
Bond lengths in relaxed (Be,Zn)Se supercells;
A discrete set of 128 nearest-neighbor distances is artificially smeared 
for better visibility.
Top panels: bond lengths in the ``model'' supercell, with four Be atoms
in continuous Be--Se chains, and two isolated Be impurities. Distances
between different types of anions and cations are shown separately.
Bottom panels: Be--Se and Zn--Se interatomic distances in four supercells,
containing one to four Be atoms, neighboring the same Se atom.
}
\end{figure}

The diversity of local order in a mixed crystal competes with the
tendency to maintain the bond lengths, with the effect that the latter 
get a certain scattering around their mean values. This is illustrated
in Fig.~\ref{fig:lattice} for the Be$_6$Zn$_{26}$Se$_{32}$ supercell, 
and in more detail for this and other ``large'' supercells --
in Fig.~\ref{fig:bonds}. A discrete ``spectrum'' of, in total, 128 bond lengths 
in each fully relaxed supercell is artificially broadened there 
for better visibility.
We start our discussion from the BeZn$_{31}$Se$_{32}$ supercell.
Four Se neighbors experience an inward relaxation to a single Be impurity,
shortening the Be--Se bonds to about 2.24 {\AA}, i.e., only slightly larger 
than in pure BeSe. Simultaneously the bonds to their Zn neighbors extend
beyond the average interatomic distance ($\sim$2.41 {\AA}) in the rest
of the predominantly-ZnSe supercell. By gradually adding more Be neighbors 
which share the same Se atom, we end up 
with a symmetric relaxation pattern in Be$_4$Zn$_{26}$Se$_{32}$
with a set of 12 contracted bonds (between Be and their outer Se neighbors)
and 4 extended bonds (from each Be to the central Se).
This is clearly seen in Fig.~\ref{fig:bonds}. 
The intermediate cases of 2 and 3 Be atoms in the supercell offer
an interpolation between the two discussed cases, introducing
a diversity in the Be--Se bond lengths depending on 
the type of second neighbors. The splitting of the Be--Se bonds into shorter
and longer ones, around the mean value of $\sim$2.24 {\AA},
leads to the diversification of the Zn--Se bonds into, correspondingly,
longer and shorter ones, as compared to the dominating bulk-ZnSe
central peak at $\sim$2.41 {\AA}.
A more complex distribution of bond lengths comes about in a supercell
which contains, along with isolated Be impurities,  
a continuous --Be--Se-- chain (Fig.~\ref{fig:bonds}, top panels). 
A Be atom substituting Zn favors an inward relaxation of neighboring Se.
Large enough around a zero-dimensional defect (isolated atom), 
such inward relaxation is even stronger around a one-dimensional defect,
such as an extended --Be--Se-- chain. The tendency for
shortening the Be--Se bond lengths should obviously exist also along the chain, 
but the actual contraction is limited by the fact that the chain is 
quasi-infinite and embedded into the ZnSe lattice, which effectively fixes
the chain's step. However, since the repeated cation-anion sequence 
in the zincblende structure is folded, the shortening of its links can be 
achieved by accommodating the interbond angles, at the price of
stretching the bonds between the in-chain Se atoms and their Zn neighbors
in the crystal (shaded area in the top right panel of Fig.~\ref{fig:bonds}).
The distribution of bond lengths between Zn cations and the off-chain Se 
anions resembles roughly that in the chain-free Be$_4$Zn$_{28}$Se$_{32}$
supercell.

\section[#5]{Force constants}
\label{sec:forces}
\begin{figure}
\centerline{\epsfig{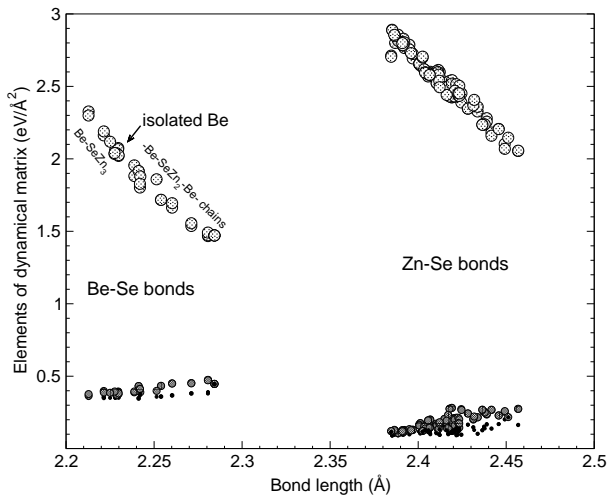}}
\caption{\label{fig:forceNN}
Diagonal elements of force constant matrix between nearest neighbors.
See text for details.
}
\end{figure}

We look now at how the variations in bond lengths map onto changes in
the force constants. As already mentioned, the latter become accumulated
as we displace the atoms one by one 
from their equilibrium positions along three Cartesian directions 
and analyze the forces induced on all atoms of the supercell.
Specifically (in the symmetrized form), 
\begin{eqnarray}
D^{\alpha\beta}_{ij} = - \frac{1}{2} \left[
\frac{
F^{\alpha}_i(\{{\bf R}\}+d^{\beta}_j) -
F^{\alpha}_i(\{{\bf R}\}-d^{\beta}_j)
}{2d^{\beta}_j}
\right. \nonumber
\\
+
\left.
\frac{
F^{\beta}_j(\{{\bf R}\}+d^{\alpha}_i) -
F^{\beta}_j(\{{\bf R}\}-d^{\alpha}_i)
}{2d^{\alpha}_i}
\right]
\label{eq:force}
\end{eqnarray}
where $F^{\alpha}_i$ is the force on atom $\alpha$ in the direction $i$,
and $\{{\bf R}\}$+$d^{\beta}_j$ means that of all atoms, only the atom $\beta$
is displaced along $j$ from its equilibrium position, by (in our case)
$d$=0.03 Bohr. With all elements of $D^{\alpha\beta}_{ij}$ recovered, 
the solution of dynamical equation
\begin{equation}
\sum_{\beta,k} \left[\,
\omega^2 \delta_{\alpha\beta}\delta_{ik} -
\frac{D^{\alpha\beta}_{ij}}{\sqrt{M_{\alpha}M_{\beta}}}\right]
A^{\beta}_k
= 0
\label{eq:dynam}
\end{equation}
yields zone-center phonon frequencies $\omega$ and eigenvectors 
$A^{\alpha}_i$ of the supercell.
Before turning to the analysis of phonons, we discuss briefly the force
constants $D^{\alpha\beta}_{ij}$. For each atomic pair $\{\alpha,\beta\}$, 
they make a 3$\times$3 matrix which can be diagonalized. 
The diagonal elements are shown in Fig.~\ref{fig:forceNN} in dependency 
on the corresponding interatomic distance. In accord with previously
discussed scattering of the Be--Se and Zn--Se bond lengths, we have now 
a scattering of force constants.
The major elements (of diagonalized 3$\times$3 matrix) show a remarkably
pronounced linear variation with the bond length. Two minor elements
are much smaller in magnitude and roughly bondlength independent.
One can interpret the smallness of minor elements 
of $D^{\alpha\beta}$ as a measure of the covalency of corresponding bonds.
Indeed, in case of a purely ionic bonding only the central motion
of atoms would bring about a change in the force; the tangential component
makes a non-zero contribution to the force constant only if
tangential displacement induces a re-distribution of charge density, 
meaning that the covalency of the bond is not negligible.
The contrast between strongly anisotropic and hence essentially ionic
force constants for the Zn--Se bonds and much more covalent Be--Se
interaction is clearly seen in Fig.~\ref{fig:forceNN}. It is remarkable
that the central Be--Se interaction is, on the average, smaller 
than the Zn--Se one. Yet, we have seen that the BeSe crystal possesses
larger bulk modulus and a higher elasticity parameter in the 
$\Gamma$-TO vibration. This can only be explained by a considerable
contribution of non-central forces (minor elements of the force constants)
in the shaping of such elastic parameters. In other words, 
whereas in ZnSe the stretching of Zn--Se bonds plays a dominant role 
in the elasticity, in BeSe the bending of bond angles is nearly as important. 

\begin{figure}
\centerline{\epsfig{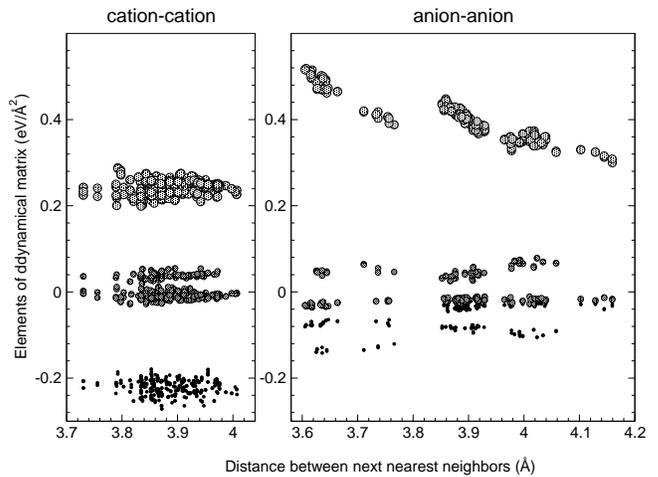}}
\caption{\label{fig:forceNNN}
Diagonal elements of force constant matrix between next nearest neighbors.
Left panel: (Be,Zn)--(Be,Zn); right panel: Se--Se.
}
\end{figure}

We find very remarkable a pronounced linear dependence of the force 
constants on the interatomic distance, in apparent indifference
to other aspects of the local order. The obtained dependence can be
parametrized and used for fast lattice-dynamics simulations in
much larger supercells. The results for similarly diagonalized
second-nearest neighbors interaction are shown in Fig.~\ref{fig:forceNNN}.
The scattering of the Se--Se distances is twice as large as that of 
cation-cation ones, and shows a clear splitting into two groups,
similar to that reported in Ref.~\onlinecite{APL82-1057} for the
As--As pair distribution function in (Ga,In)As.
This means that Be substitutes Zn almost at their original
(non-displaced) positions, i.e., the cation sublattice remains rather rigid,
whereas the Se atoms undergo large displacements,
depending on their local environment.
We see a pronouncedly different behavior of cation-cation
and anion-anion force constants: the former have two almost symmetric
(positive/negative) diagonal elements, with the third being nearly zero,
independently on both type of pair (Be--Be, Be--Zn or Zn--Zn)
and on the distance. The anion-anion coupling has an appreciable 
central-force element, decreasing with the distance, and two much smaller 
tangential elements. 
The variation of the major element with distance is (at least in part) 
related to the fact that on the left hand side of the plot we deal mostly with
Se--Be--Se connections, in which the central Se--Se interaction reflects the
effect of bending the strongly covalent Be--Se bonds; on the right hand site,
the Se--Zn--Se connections offer a much weaker resistance to such bending,
due to a higher bond ionicity.

\section[#6]{Phonons}
\label{sec:phonons}
\begin{figure}
\centerline{\epsfig{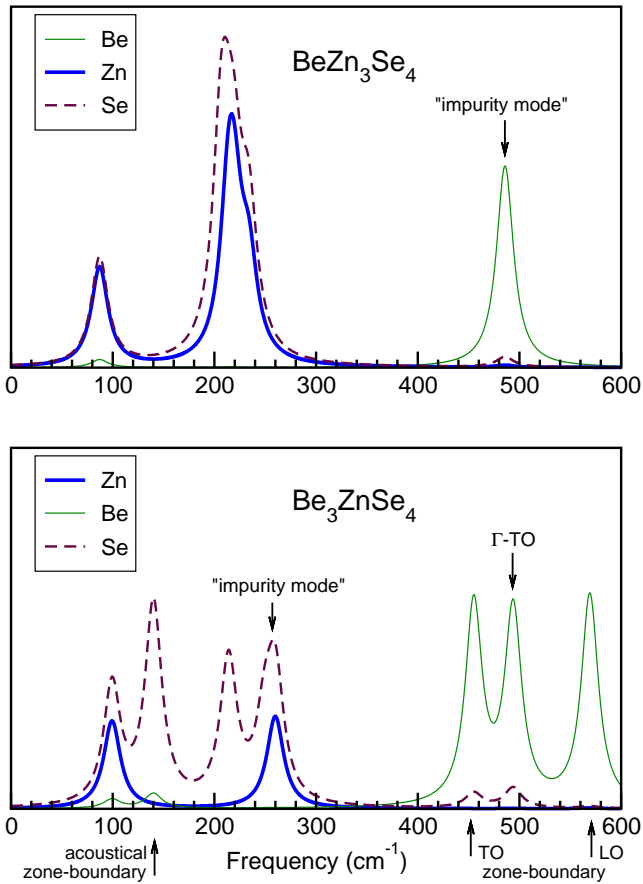}}
\caption{\label{fig:4x-PhDOS}
Phonon density of states for the BeZn$_3$Se$_4$ and Be$_3$ZnSe$_4$ supercells,
calculated for $q$=0 of the supercell and broadened by 10 cm$^{-1}$.
}
\end{figure}

Since we deal with relatively large ``disordered'' supercells with no internal
(short-scale) periodicity, and no clear-cut phonon dispersion 
exists in such systems, it does not make sense to Fourier-transform 
the force constants (\ref{eq:force}) prior to solving the dynamical
equation (\ref{eq:dynam}). However, we explain below how to extract 
some $q$-resolved information about lattice vibrations.
The primary characteristic for our discussion
will be the phonon density of states (PhDOS), resolved when necessary over 
freely chosen groups $\aleph$ of atoms $\alpha$:
\begin{equation}
I_{\aleph}(\omega) = \sum_{\alpha \in \aleph}\sum_i
\left|A^{\alpha}_i(\omega)\right|^2\,.
\label{eq:phDOS}
\end{equation}
This is a discrete spectrum of $(3N-3)$ lines (for $N$ atoms in the supercell,
with acoustic modes removed) of different intensity, which is in the
following figures broadened with the halfwidth parameter of 10 cm$^{-1}$,
for better visibility.
The vibration modes obtained from the solution of Eq.~(\ref{eq:dynam})
correspond to the zone-center of the supercell in question, but they
reflect different vibration patterns, also those of non-zone-center
character, with respect to the underlying zincblende lattice.
Let us discuss this for 8-atom supercells, whose calculated PhDOS 
is shown in Fig.~\ref{fig:4x-PhDOS}.

The $q$=0 vibration modes in a supercell include the ``true'' 
zone-center phonon (relative to the basic zincblende cell), as well as
zone-boundary phonons backfolded to a smaller BZ of the superstructure.
It is relatively easy to distinguish them by their 
eigenvectors; certain modes are labeled in Fig.~\ref{fig:4x-PhDOS}.
Most obviously, the acoustical branch hits the BZ boundary at
87 cm$^{-1}$ (Zn-rich system) or 140 cm$^{-1}$ (Be-rich system).
As ranges of frequencies of Be-related and Zn-related optical modes
are well separated, due to a large difference in masses,
it is straightforward to recognize the BeSe-type $\Gamma$-TO mode
at 494 cm$^{-1}$ (lower panel of Fig.~\ref{fig:4x-PhDOS}), the backfold
of the optical branch at the zone boundary at 450 cm$^{-1}$, 
and the backfold of the zone-boundary longitudinal phonon at 560 cm$^{-1}$ --
all in good agreement with the linear-response phonon dispersions 
calculated for BeSe by Tutuncu \emph{et al.}\cite{BeSe-Tutuncu}.
The zone-center LO mode is not present, because we do not include
macroscopic electric field in crystal in our calculation.
In the Zn-rich crystal, the Be-related TO mode (at 487 cm$^{-1}$) behaves 
like an impurity mode and exhibits no dispersion.
Whereas the Zn-related mode is similarly dispersionless (at 260 cm$^{-1}$)
in the Be-rich system, in BeZn$_3$Se$_4$ the ZnSe-related TO branch
shows dispersion, manifested by backfolding of zone-boundary phonon
and hence additional structure at 200--240 cm$^{-1}$, in agreement
with experiments and calculations of Refs.~\onlinecite{PhL36A-376,PSSB229-563}.

Even as these 8-atom supercells may be too simplistic to imitate short-range
order effects in real quasibinary alloy, they exhibit important features
which also persist in more sophisticated representative structures:
a difference between dispersionless modes due to ``isolated'' impurities
and dispersing modes due to continuous connected chains; 
a more pronounced dispersion in BeSe-related modes than in much softer
ZnSe-related ones.

\begin{figure}
\centerline{\epsfig{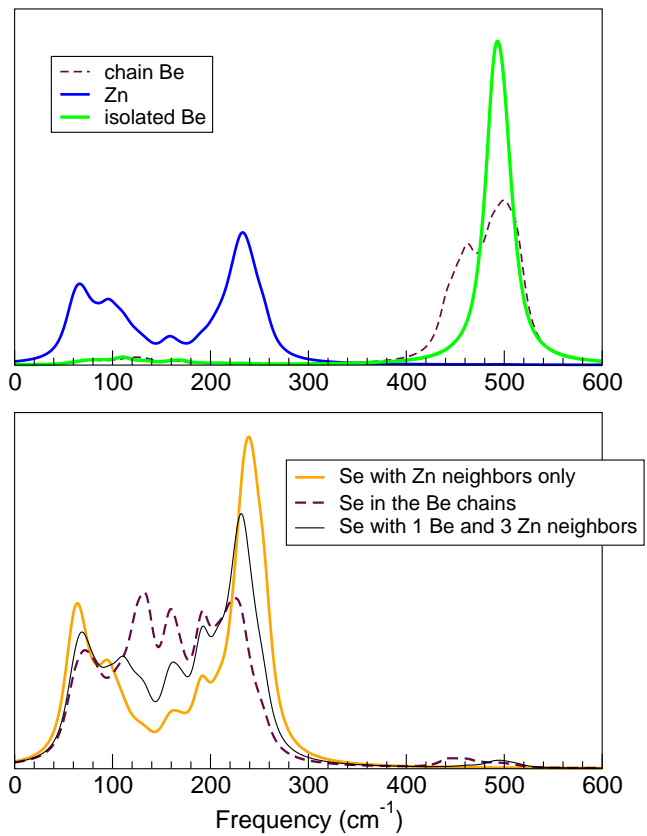}}
\caption{\label{fig:32x-PhDOS}
Phonon density of states for the Be$_6$Zn$_{26}$Se$_{32}$ supercell, 
resolved over different groups of atoms,
calculated for $q$=0 of the supercell and broadened by 10 cm$^{-1}$.
The vertical scaling for different groups is arbitrary.
}
\end{figure}

We turn next to the discussion of phonons in a larger Be$_6$Zn$_{26}$Se$_{32}$
supercell. There are now more lines in the spectrum, and, as the Brillouin zone
(BZ) becomes smaller, the $\Gamma$ modes of the supercell offer 
a sampling over  more non-zone-center modes in a zincblende lattice. 
This leads in more features and more ``filled'' form of the PhDOS 
(Fig.~\ref{fig:32x-PhDOS}).

The vibration spectra of both pure systems ZnSe and BeSe are shaped
by broad acoustical branches and more narrow optical ones.
In ZnSe they nearly overlap, resulting in a characteristic two-peaked
structure of the PhDOS -- see, e.g., Hennion \emph{et al.}\cite{PhL36A-376}.
Our low-frequency (ZnSe-related, up to 300 cm$^{-1}$) part 
of the supercell PhDOS provides a fair agreement with this known
result for pure ZnSe.
The spectrum of BeSe has a large separation between its acoustical
and optical parts. The former covers the whole
region of the phonon spectrum of ZnSe. Therefore, in a mixed crystal,
the partial PhDOS of Se atoms which have four Zn neighbors
essentially repeats the shape of the Zn PhDOS, whereas those Se atoms
with one or two Be neighbors participate in vibrations at
100 -- 200 cm$^{-1}$, i.e., in the ``pseudogap'' region of ZnSe. 
These modes have mostly acoustical character.

We consider now more attentively the high-frequency optical modes 
with high participation of Be vibrations. 
The corresponding PhDOS (Fig.~\ref{fig:32x-PhDOS}, top panel) is separated 
into contributions from (two) ``isolated Be'' atoms and those (four)
in the infinite chain. The ``isolated Be'' contribution is, expectedly,
a single narrow line, coming about from several nearly degenerate vibration
frequencies. The vibrations of Be in the chains are more diversified:
the characteristic frequencies can be found below, within and above
the ``isolated Be'' peak. The most remarkable is a clear split off
at a soft side, which makes a distinguished peak in the vibration spectrum,
similar to that experimentally observed in samples with a presumed
percolation on the Be sublattice\cite{APL77-519,PRB65-035213}.
The previous discussion on the distribution of bondlengths, and
the force constants' dependence on the latter, helps us to understand
this behavior. The bonds between the chain Be atoms and 
\emph{out-of-chain} Se are shorter that those involving isolated Be,
and corresponding force constants are larger. On the contrary,
the Be--Se bonds \emph{along the chain} are less contracted; the longer
bond length implies smaller force constant hence lower vibration frequency.
Our explanation of the experimentally observed anomalous (split-off) 
Be mode is, therefore, the following. As the concentration of Be in ZnSe 
reaches the percolation limit and infinite chains are built, the Be--Se
bonds in these chains are forced to become longer that around isolated Be
impurities, or in short chain fragments. This immediately softens the
vibration modes which involve such extended bonds.

\begin{figure}
\centerline{\epsfig{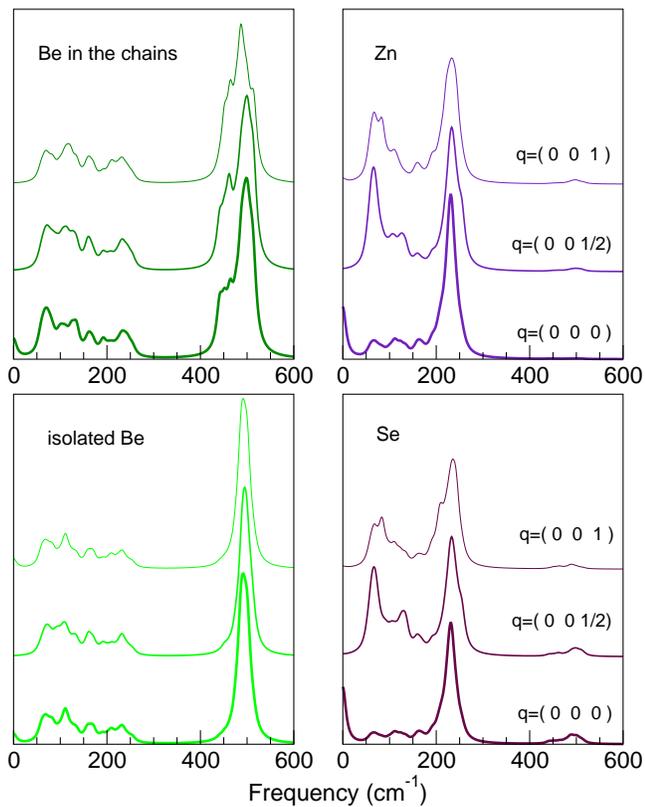}}
\caption{\label{fig:q-PhDOS}
Phonon spectral function in the Be$_6$Zn$_{26}$Se$_{32}$ supercell, 
corresponding to three values of $q$ of the basic zincblende lattice,
broadened by 10 cm$^{-1}$.
}
\end{figure}

In order to extend this interpretation over experimentally measured
Raman lines, which are known to scan primarily the zone-center phonons,
one needs yet to extract a wavevector-resolved information about calculated 
vibrations. As the supercell size increased, the effective sampling over the
Brillouin zone (BZ) of the  basic zincblende lattice became more fine.
In particular, one has not only $X$, $K$ and $L$ zone-boundary phonons
sampled, as was the case for the 8-atom cells, but also $q$ points halfway
to them from the zone center. We mentioned already that the lack 
of translation symmetry in the mixed BeSe--ZnSe supercell does not allow 
to Fourier transform the dynamical matrix and arrive at neat phonon 
dispersion curves. However, one can project the dispersion patterns, 
corresponding to different modes of vibration, onto the plane wave with 
different $q$ values and in such way define $q$-resolved contributions 
to the phonon DOS:
\begin{equation}
I_{\aleph}(\omega,{\bf q}) = \sum_{\alpha \in \aleph}\sum_i
\left|A^{\alpha}_i(\omega)\exp(i{\bf q}{\bf R}_{\alpha})\right|^2\,.
\label{eq:phon_q}
\end{equation}
Since eigenvectors correspond to $q$=0 of the supercell and hence are real,
the phase of the plane wave has no importance. 
One can view $I_{\aleph}(\omega,{\bf q})$ as spectral function, which
in principle would show how the phonon dispersion bands of pure constituents
get overlapped, distorted and smeared in a mixed crystal. Unfortunately,
the results obtained for our -- yet relatively small -- 64-atom supercell
do not allow to see anything resembling a continuous displacement of intensity
maxima with $q$\cite{note_on_q}.
Nonetheless, we present in Fig.~\ref{fig:q-PhDOS} the projected
PhDOS for three $q$ values along the (001) direction. 
The ZnSe-type vibrations make clearly visible the displacement 
of two acoustical bands, known, e.g., from the measured phonon 
dispersion curves\cite{PhL36A-376}, as the projection wavevector changes 
from the zone-center via midpoint to the zone boundary. 

\begin{figure*}[t!]
\centerline{\epsfig{figure=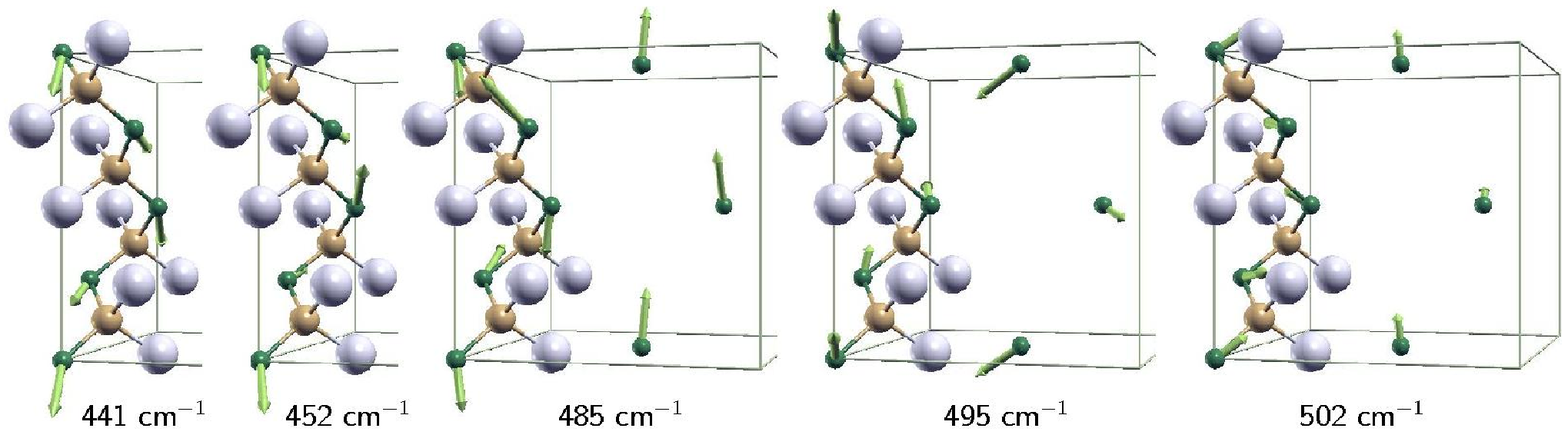,width=18.0cm}}
\caption{\label{fig:modes}
Vibration patterns of five selected modes with substantial contribution
of Be atoms; the relative displacements of the latter are shown by arrows.
Only some atoms of the Be$_6$Zn$_{26}$Se$_{32}$ supercell
(Fig.~\ref{fig:all_scells}, right panel) are shown.
The figure was created with the XCrySDen software \cite{xcrysden}.
The first two modes affect exclusively the continuous --Be--Se-- chains;
the last three contain also comparable contribution of
impurity Be atoms. See text for details.
}
\end{figure*}

For the Be-related modes, because of small amount of Be atoms in the supercell,
the dispersion is generally less pronounced, yet noticeable in the (001) 
direction, which is the general direction of the chains. 
Being almost non-interacting linear defects,
the chains do not give rise to noticeable dispersion in perpendicular 
directions. The three acoustical modes, removed in the previous
PhDOS figures, are retained in Fig.~\ref{fig:q-PhDOS},
in order to show how the spectral weight of the $\omega$=0 modes
disappears for $q$ values away from zero. A notable difference is
the PhDOS of isolated Be, which is nearly identical for $q$=0 and
$q$=(001). This is because the supercell contains two isolated Be
atoms. They interact only weakly but create a superstructure that 
effectively halves the BZ in the $z$ direction.

The $q$-resolving of the PhDOS singles out the modes which
are likely to dominate the Raman spectra. The separation
into $q$=0 and non-$q$=0 modes is however not very clear-cut --
in part due to physical reasons (mixed crystal, loss of periodicity)
and in part due to technical limitations (small size of supercell
and hence non-vanishing $q$=0 projection of many modes due to
numerical noise). One can expect better contrast in selecting $q$=0
modes in a larger supercell, due to better sampling.
Nevertheless, it is clear that the separation into ``isolated Be''
and ``Be in the chains'' modes occurs not only on the average
over the BZ, but is actually even more pronounced in the zone-center modes. 
Therefore, our explanation of anomalous Raman lines will hold.

Most of the modes are of very mixed character, and involve
stretching and bending of bonds in a complicated manner. However, we selected
several ones, involving the --Be--Se-- chains,
whose vibration pattern can be described with relative
simplicity. Two of these modes (441 and 452 cm$^{-1}$)
lie in the ``anomalous'' Raman line, where the vibrations of isolated
Be atoms are negligible. Three other selected modes,
at 485, 495 and 502 cm$^{-1}$, involve simultaneous vibration of all
Be atoms, in the chains as well as isolated ones.
Fig.~\ref{fig:modes} gives a snapshot of these five modes. 
Displacement patterns in some other modes are shown in Fig.~5 
of Ref.~\onlinecite{Dui03-ZBS}.
The vibrations of predominantly ``impurity Be'' character,
leaving the chain Be atoms silent, occur at nearly 490 cm$^{-1}$,
in between the third and the fourth of depicted modes.

The most general observation is that the ``anomalous'' modes 
with reduced frequencies involve nearly in-plane displacement
of Be with its own equilibrium position and its two Se neighbors.
Such motion stretches or shortens the Be--Se
bonds but roughly preserves the Be--Se--Be angles. 
The 441 cm$^{-1}$ mode is a wave of correlated Be displacement along
the chain, periodic with the chain step. This is a predominantly $\Gamma$ mode.
The 452 cm$^{-1}$ mode has similarly ``longitudinal'' character but
the doubled translation length: two Be atoms approach, or depart from,
their common Se neighbor symmetrically. Consequently this is not 
a clear zone-center vibration, however due to the curvature of the chain and 
a complexity of detailed displacement pattern it provides a ${\bf q}$=0 
contribution in the analysis according to Eq.~(\ref{eq:phon_q}).

The harder modes are related to out-of-plane Be vibrations and hence bending of
covalent Be--Se bonds. Somehow simplifying, in the ``anomalous modes''
there are primarily central Be--Se forces at work, so the frequency is lower
than that involving an isolated Be impurity -- in accord with longer
anion-cation distance along the chain, as illustrated 
by Fig.~\ref{fig:forceNN}. In the latter case, even as interatomic distances
remain the same, the off-center contributions due to the bond bending
effectively increase the force constants and harden the resulting
frequency beyond those for an isolated Be. 

\section[#7]{Conclusions}

We simulated the equilibrium structure and lattice dynamics 
from first principles in a series of supercells simulating
mixed (Be,Zn)Se crystals. We found a nearly linear dependence
of equilibrium lattice constant with composition, but a relative
independence of (average) Be--Se and Zn--Se distances on concentration.
When incorporated in a mixed crystal, the individual interatomic distances
develop a certain scattering around their mean values, maintaining
however a clear separation between the Be--Se and Zn--Se bond lengths.
Specifically, an isolated Be atom may much more efficiently shorten the bonds
to an Se neighbor whose other three bonds are Zn-terminated, than to 
an Se atom shared by other Be neighbors. In particular, a continuous 
-Be--Se- chain is caracterized by longer link lengths than the average 
Be--Se distance, for a given concentration.
We demonstrated further that this has effect on force constants and
vibration frequencies.
As follows from our direct calculation of interatomic force constants,
the principal values of the latter, taken as diagonal elements
of a 3$\times$3 matrix relating two selected atoms, show a linear
decrease as a function of cation-anion distance between nearest neighbors.
The calculations manifest a primarily ionic character of the Zn--Se
interaction and a remarkable level of covalency in the Be--Se bonds.
A decrease of force constants with larger bond distance is the
microscopic origin of the formation of a softened ``anomalous'' mode, 
which appears on the low-frequency side of BeSe-related TO mode.
Our wavevector analyzis of vibration patterns in the supercell indicates 
that the ``anomalous'' mode has a strong zone-center contribution,
that is consistent with its clear experimental observation in the Raman spectra.
We confirmed therefore on the microscopic level the early presumed relation 
between the onset of Be percolation threshold on the cation sublattice
and the appearance of ``anomalous'' mode, including specific vibrations
of continuous --Be--Se-- chains transversing the crystal.

\section*{Acknowledgments}
A.V.P. thanks the Universit\'e Metz for invitation for a research stay there
and the hospitality of the Institut de Physique.
Useful comments by N.~E.~Christensen, discussions with V.~Vikhnin 
and travel support from the NATO project CLG 980378 are greatly appreciated.


\begin{thebibliography}{37}
\bibitem[{not()}]{note_on_q}
\bibinfo{note}{The reason is that only limited number of vibration frequencies
  is represented in the supecell, and only the weight of different modes but
  not their frequencies may change in the projection on different $q$. For
  instance, there are not enough modes corresponding to the gradual rise of
  acoustical branches towards the zone boundary. As a consequence, the phonon
  DOS plotted in the ($q,\omega$)-domain shows a step-like rather than smooth
  change.}
\end{thebibliography}

\end{document}